\documentclass{ws-jnopm}

\usepackage{graphicx}

\begin{document}
 
%\markboth{Authors' Names}{Paper's Title}

%%%%%%%%%%%%%%%%%%% Publisher's Area please ignore %%%%%%%%%%%%%%%%%%%%%%%
%\catchline{}{}{}{}{}
%%%%%%%%%%%%%%%%%%%%%%%%%%%%%%%%%%%%%%%%%%%%%%%%%%%%%%%%%%%%%%%%%%%%%%%%%%

\title{
DYNAMICS OF FERROELECTRIC NANO CLUSTER IN BaTiO$_3$\\
OBSERVED AS A REAL TIME CORRELATION BETWEEN TWO\\
SOFT X-RAY LASER PULSES}

\author{KAI JI} 
\address{
Solid State Theory Division, Institute of Materials Structure Science,\\
KEK, Graduate University for Advanced Studies, and CREST JST\\
Oho 1-1, Tsukuba, Ibaraki 305-0801, Japan\\
jikai@post.kek.jp
}

\author{KAZUMICHI NAMIKAWA} 
\address{
Department of Physics, Tokyo Gakugei University\\
Nukuikita 4-1-1, Koganei, Tokyo 184-8501, Japan
}

\author{HANG ZHENG} 
\address{
Department of Physics, Shanghai Jiao Tong University\\
Shanghai 200030, China
}

\author{KEIICHIRO NASU} 
\address{
Solid State Theory Division, Institute of Materials Structure Science,\\
KEK, Graduate University for Advanced Studies, and CREST JST\\
Oho 1-1, Tsukuba, Ibaraki 305-0801, Japan
}

\maketitle

\begin{abstract}
We carry out a theoretical investigation to clarify the dynamic property
of photo-created nano-sized ferroelectric cluster observed in the paraelectric
BaTiO$_3$ as a real time correlation of speckle pattern between two soft
X-ray laser pulses, at just above the paraelectric-ferroelectric phase
transition temperature.
Based on a model with coupled soft X-ray photon and ferroelectric phonon
mode, we study the time dependence of scattering probability by using
a perturbative expansion approach.
The cluster-associated phonon softening as well as central peak effects
are well reproduced in the phonon spectral function via quantum Monte
Carlo simulation.
Besides, it is found that the time dependence of speckle correlation is
determined by the relaxation dynamics of ferroelectric clusters.
Near the transition point, cluster excitation is stable, leading to a
long relaxation time.
While, at high temperature, cluster structure is subject to the thermal
fluctuation, ending up with a short relaxation time.
\end{abstract}


\section{Introduction}

Ferroelectricity refers to the ability of developing a spontaneous polarization
(electric dipole moment) on decreasing temperature below Curie point ($T_c$).
As a prototype of the ferroelectric perovskite compounds, barium titanate
(BaTiO$_3$) undergoes a transition from paraelectric cubic to ferroelectric
tetragonal phase at $T_c$=395 K.
In addition to its extensive application in the technology such as non-volatile
memory devices, capacitor applications and piezoelectric transducers due
to its high dielectric constant and switchable spontaneous
polarization,\cite{po98} there is also an enduring interest in understanding
the mechanism of paraelectric-ferroelectric phase transition in it.
Although it is generally considered that the transition might be a classic
displacive soft-mode type driven by the anharmonic lattice
dynamics,\cite{ha71,mi76} recent studies have suggested an existence
of order-disorder instability competing with the displacive
transition\cite{za03,vo07}, making this issue still controversial to date.

Direct observation on creation and evolution of ferroelectric cluster
around $T_c$ is of crucial importance for clarifying the nature of phase
transition.
However, conventional techniques, like neutron scattering,\cite{ya69}
polarizing optical microscopy\cite{mu96} and scanning probe
microscopy,\cite{pa98} can hardly detect the ultrafast transient status
of dipole clusters.
In contrast, diffraction speckle patterns of BaTiO$_3$ crystal measured
by the soft X-ray laser has turned out to be an efficient way to elucidate
the dynamics of domain structure.$^{9-11}$%\cite{go07,ta02,ta04}

Very recently, Namikawa\cite{na08} study the nano-sized
polarization clusters in BaTiO$_3$ at just above $T_c$ by the plasma-based
X-ray laser speckle measurement.
In this experiment, two consecutive coherent X-ray laser pulses of 160
\AA\ with an adjustable time difference are generated by the Michelson
type beam splitter.
After the photo excitation by the first pulse, ferroelectric clusters
of nano scale are created in the paraelectric BaTiO$_3$ and tends to be
smeared out gradually due to the thermal fluctuation.
This relaxation of cluster structure thus can be recorded in the speckle
patterns of the second pulse as a function of its delay time from the
first pulse, which is known as the speckle intensity correlation function.
It has been found that near the $T_c$, this correlation follows an exponential
decay process, {\it i.e.}, the intensity of speckle pattern declines as
the delay time increases.
Moreover, the decay rate also decreases upon approaching $T_c$, indicating
a critical slowing down of the dipole relaxation time.
Hence, by measuring the correlation between two soft X-ray laser pulses,
the real time relaxation dynamics of polarization clusters in BaTiO$_3$
is clearly represented.

In this work, we examine the above-mentioned novel behaviors of ferroelectric
cluster observed by Namikawa from a theoretical point of view, aiming to
provide a basis for understanding the critical nature of BaTiO$_3$.
Our calculation confirms that the relaxation of photo-created nano clusters
plays an essential role in determining the delay time dependence of
speckle correlation function.
The remaining of present paper is organized as follows.
In Sec. 2, the model Hamiltonian and theoretical treatment are put forward.
In Sec. 3, our numerical results are discussed in comparison with the
experimental discoveries.
In Sec. 4, a summary with conclusion is presented finally.

\section{Theoretical model and method}

In order to describe the optical response of BaTiO$_3$ due to X-ray scattering,
we design a theoretical model to incorporate the radiation field and
ferroelectric dipole correlation as well as a weak interplay between them.
As mentioned above, the phase transition in BaTiO$_3$ is likely to be
an admixture of displacive and order-disorder types.
For this sake, we adopt a Krumhansl-Schrieffer type phonon mode\cite{kr75}
to characterize the ferroelectric anharmonicity in BaTiO$_3$, for this
model is suitable for describing the crossover between displacive and
order-disorder phase transitions.\cite{ru01}
In this scenario, our model Hamiltonian reads ($\hbar = 1$ and $k_B = 1$
throughout this paper):
\begin{eqnarray}
H &=& H_p + H_f + H_{pf} ,
    \nonumber\\
H_p &=& \sum_k \Omega_k a_k^{\dag} a_k,
    \nonumber\\
H_f &=& {\omega_0 \over 2} \sum_l \left(
    -{{\partial}^2 \over \partial Q_l^2} + Q_l^2 - c_4 Q_l^4 + c_6 Q_l^6
    \right) -{\omega_0 d_2 \over 2} \sum_{<l,l'>} Q_l Q_{l'}
    \nonumber\\
H_{pf} &=& {V \over N} \sum_{q,q',k} a_{k + {q \over 2}}^{\dag}
    a_{k - {q \over 2}} Q_{q' - {q \over 2}} Q_{-q' - {q \over 2}} ,
\end{eqnarray}
where $a_k^{\dag}$ ($a_k$) is the creation (annihilation) operator of
a photon with a wave vector $k$ and an energy $\Omega_k$.
$Q_l$ is the coordinate operator for ionic displacement along the easy
axis of BaTiO$_3$ at the site $l$ with a dipole oscillatory frequency
$\omega_0$.
$Q_q$ ($\equiv N^{-1/2} \sum_l e^{-i q l} Q_l$) is its Fourier component
for wave vector $q$.
Without losing generalitivity, here we use a simple cubic lattice, and
the total number of lattice site is $N$.
$H_f$ is the extended Krumhansl-Schrieffer model for phonon, where a
sixth order term of $Q_l$ is introduced to produce a triple-well
potential, because the phase transition of BaTiO$_3$ is first order.
Thus, the lattice dynamic properties are determined by the frequency
$\omega_0$, temperature $T$, and dimensionless parameters $c_4$, $c_6$, $d_2$.
$H_{pf}$ corresponds to a Raman type bi-linear coupling between the
ferroelectric phonon mode and photons as the phonon mode has an even parity
in the present case, and $V$ is the coupling strength.

Since there are two photons involved in the scattering with crystal, the
photon-phonon scattering probability can be written as,
\begin{eqnarray}
P(t) &=& \sum_{k_1, k'_1} \langle \langle
    a_{k_0}(0) a_{k_1}^{\dag}(\Delta) a_{k_0}(t) a_{k'_1}^{\dag}(\Delta + t)
    \nonumber\\
& & \times
    a_{k'_1}(\Delta + t) a_{k_0}^{\dag}(t) a_{k_1}(\Delta) a_{k_0}^{\dag}(0)
    \rangle \rangle ,
\end{eqnarray}
where,
\begin{eqnarray}
\langle \langle \cdots \rangle \rangle =
\mbox{Tr} (e^{- \beta H} \cdots) / (e^{- \beta H}),
\end{eqnarray}
means the expectation, $\beta$ ($\equiv 1/T$) is the inverse temperature,
and the time dependent operator $O(t)$ is defined in the Heisenberg
representation,
\begin{eqnarray}
O (t) = e^{i t H} O e^{- i t H} .
\end{eqnarray}
Here, $t$ denotes the time difference between two incident X-ray laser
photons, and $k_0$ is the wave vector of them.
After a small time interval $\Delta$, these photons are scattered out,
respectively.
$k_1$ and $k'_1$ are the wave vectors of the first and second outgoing photon.

Regarding $H_{pf}$ as a perturbation, we separate Hamiltonian of Eq. (2.1) as,
\begin{eqnarray}
H = H_0 + H_{pf} ,
\end{eqnarray}
where
\begin{eqnarray}
H_0 = H_p + H_f ,
\end{eqnarray}
is treated as the unperturbed Hamiltonian.
By expanding the time evolution operator in Eq. (2.4) with respect to $H_{pf}$,
\begin{eqnarray}
e^{- i t H} \rightarrow e^{- i t H_0} \left[
    1 - i \int_0^t d \tau \hat{H}_{pf} (\tau) + \cdots \right] ,
\end{eqnarray}
we find that the lowest order terms which directly depend on $t$ are of
fourth order,
\begin{eqnarray}
P(t) & \rightarrow & \int_0^{\Delta} d \tau_1 \int_0^{\Delta} d \tau_2
    \int_0^{\Delta} d \tau'_1 \int_0^{\Delta} d \tau'_2
    \sum_{k_1, k'_1} \langle \langle
    a_{k_0} \hat{H}_{pf} (\tau'_1) e^{i \Delta H_0} a_{k_1}^{\dag}
    \nonumber\\
& & \times e^{i (t - \Delta) H_f} a_{k_0} \hat{H}_{pf} (\tau'_2)
    e^{i \Delta H_0} a_{k'_1}^{\dag} a_{k'_1} e^{- i \Delta H_0} \hat{H}_{pf}
    (\tau_2) a_{k_0}^{\dag} e^{- i (t - \Delta) H_f}
    \nonumber\\
& & \times a_{k_1} e^{- i \Delta H_0} \hat{H}_{pf} (\tau_1) a_{k_0}^{\dag}
    \rangle \rangle ,
\end{eqnarray}
where the operators with carets are defined in the interaction representation,
\begin{eqnarray}
\hat{O} (\tau) = e^{i \tau H_0} O e^{- i \tau H_0} .
\end{eqnarray}

\begin{figure}
\begin{center}
\includegraphics{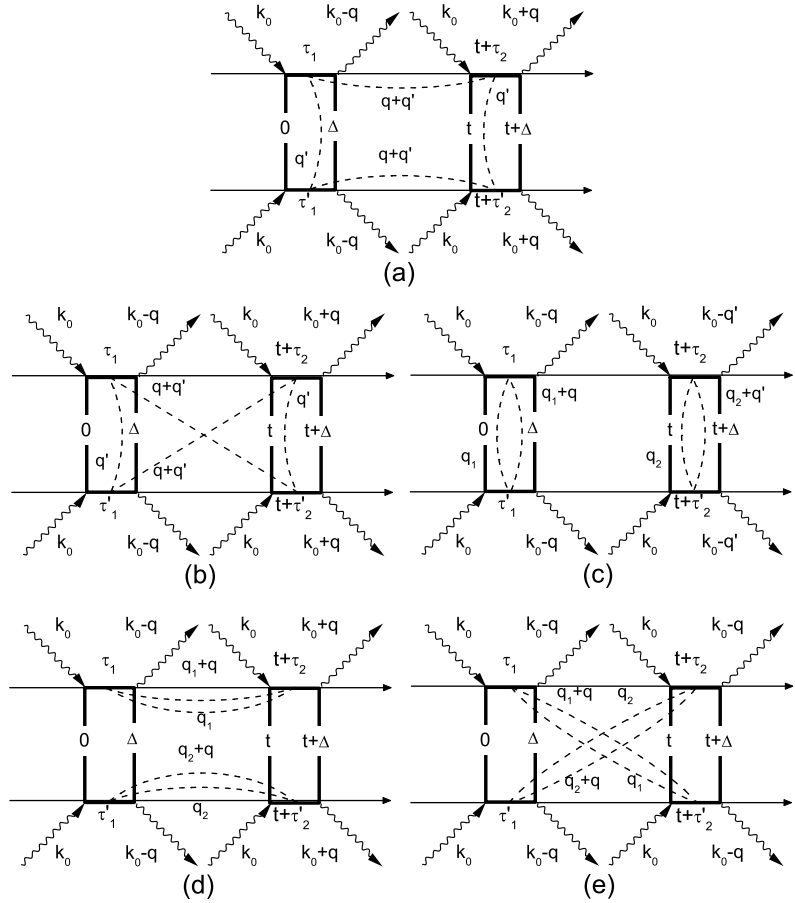}
\end{center}
\caption{
Diagrammatic analysis for scattering process of photon with electric dipole
moment (phonon).
The photons and phonons are denoted by the wavy and dashed lines, respectively.
In each diagram, the upper and lower horizontal time lines represent the
bra and ket vectors, respectively.
}
\end{figure}

In Fig. 1, we present a diagrammatic analysis for the phonon-coupled
scattering process corresponding to Eq. (2.8).
In each diagram, the photons (phonons) are depicted by the wavy (dashed)
lines, and the upper (lower) horizontal time lines stand for the bra (ket)
vectors.\cite{na94}
Diagram (a) illustrates the changes of photon energy and wave number due
to phonon emission and absorption, which corresponds to the well-known
Stokes and anti-Stokes Raman scattering.
Whereas, diagrams (b)-(e) represent the exchange, side band, rapid dumping
and rapid exchange effects, respectively.

Obviously, diagram (c) bears no $t$-dependence, while diagrams (d) and
(e) only contribute a substantial reduction to the time correlation between
two laser pulses because of the duality of phonon interchange.
Accordingly, the $t$-dependence is mainly determined by the diagrams
(a) and (b).
Allowing for this, the probability of interest can be written in the form as,
\begin{eqnarray}
P (t) &=& \int_0^{\Delta} d \tau_1 \int_0^{\Delta} d \tau_2
    \int_0^{\Delta} d \tau'_1 \int_0^{\Delta} d \tau'_2
    {2V^4 \over N^4} \sum_{q, q'}
    \langle \langle
    a_{k_0} e^{i \tau'_1 H_p} a_{k_0}^{\dag} a_{k_0 - q}
    \nonumber\\
& & \times e^{-i (\tau'_1 - \Delta) H_p} a_{k_0 - q}^{\dag}
    a_{k_0} e^{i \tau'_2 H_p} a_{k_0}^{\dag} a_{k_0 + q}
    e^{-i (\tau'_2 - \Delta) H_p} a_{k_0 + q}^{\dag} a_{k_0 + q}
    \nonumber\\
& & \times
    e^{i (\tau_2 - \Delta) H_p} a_{k_0 + q}^{\dag} a_{k_0}
    e^{-i \tau_2 H_p} a_{k_0}^{\dag} a_{k_0 - q}
    e^{i (\tau_1 - \Delta) H_p} a_{k_0 - q}^{\dag}
    a_{k_0} e^{-i \tau_1 H_p}
    \nonumber\\
& & \times a_{k_0}^{\dag} \rangle \rangle
    \langle \langle \hat{Q}_{q'} (\tau'_1) \hat{Q}_{q-q'} (\tau'_1)
    \hat{Q}_{-q+q'} (t + \tau'_2) \hat{Q}_{-q'} (t + \tau'_2)
    \hat{Q}_{q'} (t + \tau_2)
    \nonumber\\
& & \times \hat{Q}_{q-q'} (t + \tau_2)
    \hat{Q}_{-q+q'} (\tau_1) \hat{Q}_{-q'} (\tau_1)
    \rangle \rangle .
\end{eqnarray}
Here, we note that the photon and phonon parts have been decoupled, and
it can been easily seen that the origin of the $t$-dependence is nothing
but the phonon (dipole) correlation.

In the case of forward X-ray scattering, we have
$k_0$$\approx$$k_1$$\approx$$k'_1$, thus the normalized probability can
be further simplified as,
\begin{eqnarray}
\frac{P(t)}{P(0)} =
    \frac{\sum_{q,q'} | \langle \langle Q_q^2 \rangle \rangle G_{q+q'} (t) |^2}
    {\sum_{q,q'} | \langle \langle Q_q^2 Q_{q+q'}^2 \rangle \rangle |^2} ,
\end{eqnarray}
where
\begin{eqnarray}
G_q(t) = -2i \theta (t) \langle \langle \hat{Q}_q (t) \hat{Q}_{-q} (0)
    \rangle \rangle
    -2i \theta (-t) \langle \langle \hat{Q}_{-q} (0) \hat{Q}_q (t)
    \rangle \rangle ,
\end{eqnarray}
is the real time correlation function, or real time Green's function of
phonon, and $\theta (t)$ is the step function.
The Fourier component of this Green's function,
\begin{eqnarray}
G_q (\omega) = \int_{- \infty}^{\infty} dt G_q (t) e^{-i \omega t},
\end{eqnarray}
is related to the phonon spectral function [$\equiv A_q (\omega)$]
through,\cite{fe71}
\begin{eqnarray}
\mbox{Re} G_q(\omega) &=& \mbox{Re} \left[
    {1 \over 2 \pi} \int_{- \infty}^{\infty} d \omega'
    \frac{A_q (\omega)}{\omega - \omega' + i 0^{+}} \right],
    \nonumber\\
\mbox{Im} G_q(\omega) &=& - {1 \over 2} \coth \left( {1 \over 2} \beta
    \omega \right) A_q (\omega).
\end{eqnarray}
The phonon spectral function describes the response of lattice to the
external perturbation, yielding profound information about dynamic
properties of the crystal under investigation.
Once we get the spectral function, the scattering probability and
correlation function can be readily evaluated.

\section{Numerical result and discussion}

In the numerical calculation, we set phonon frequency $\omega_0$=20 meV,
and make the parameters $c_4$=0.076, $c_6$=0.00495, $d_2$=0.0967.
As indicated in Eq. (2.10), the $t$-dependence of probability is a result
of phonon correlation.
To reveal its relation with ferroelectric cluster dynamics, we start our
discussion with an inspection on the property of phonon spectral function
at the temperature regime $T$$\gtrsim$$T_c$.
To this end, we first calculate the imaginary time Matsubara Green's function
of phonon\cite{fe71} on a 12$\times$12$\times$12 cubic lattice with a
periodic boundary condition.
This is done by a path integral Monte Carlo method.\cite{ji04}
After obtaining the Matsubara function, we can derive phonon spectral
function by the analytic continuation.\cite{fe71,ji04}

\begin{figure}
\begin{center}
\includegraphics{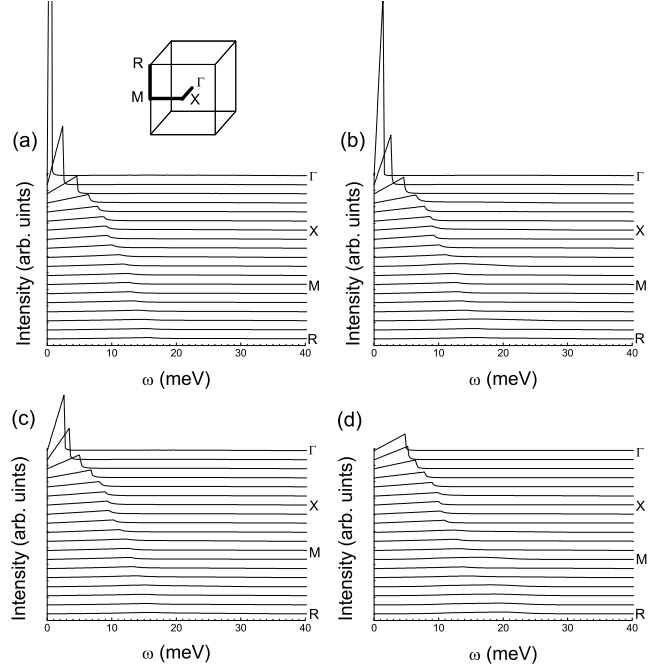}
\end{center}
\caption{
Phonon spectral function along the line $\Gamma$XMR in the Brillouin
zone at different temperatures of paraelectric phase: (a) $T$=$T_c$,
(b) $T$=1.048$T_c$, (c) $T$=1.143$T_c$ and (d) $T$=1.429$T_c$.
The inset of panel (a) shows the Brillouin zone with high symmetry lines.
}
\end{figure}

In Fig. 2, we plot the phonon spectral function in the paraelectric
phase at different temperatures: (a) $T$=$T_c$, (b) $T$=1.048$T_c$, (c)
$T$=1.143$T_c$ and (d) $T$=1.429$T_c$.
In each panel, the spectra are arranged with wave vectors along the $\Gamma$XMR
direction of Brillouin zone (see in the inset), and here $\omega$ refers
to energy.
Since the spectra are symmetric with respect to the origin $\omega$=0,
in this graph we only show the positive part of them.
When the temperature decreases toward $T_c$, as already well-known for
the displacive type phase transition, the energy of phonon peak is
gradually softened.
In addition, a so-called central peak, corresponding to the low energy
excitation of ferroelectric cluster, appears at the $\Gamma$ point.
On decreasing temperature, its intensity is dramatically enhanced, indicating
that a stable dipole correlation of long-wave-length is built up.
As mentioned above, the phonon spectral function can be understood as
the optical response of dipole to the external perturbation.
Therefore the collective excitation represented by the sharp resonant
peak in spectra is nothing but the photon-created ferroelectric cluster.

\begin{figure}
\begin{center}
\includegraphics{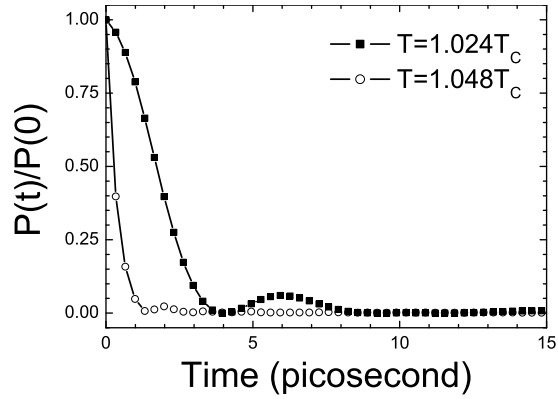}
\end{center}
\caption{
Normalized speckle scattering probability as a function of time for
paraelectric BaTiO$_3$, at $T$=1.024$T_c$ and $T$=1.048$T_c$.
}
\end{figure}

The appearance of sharp peak near $\Gamma$ point signifies that after
excitation by soft X-ray, the photon-created ferroelectric clusters
can be well preserved for a long relaxation time.
Thus, they are likely to be observed by subsequent detective photons,
resulting in a high intensity of speckle correlation.
Keeping this in mind, we move on to the results of scattering probability.
In Fig. 3, we show the normalized probability as a function of time at
temperatures $T$=1.024$T_c$ and $T$=1.048$T_c$.
From this figure, we see in both cases, the probability declines exponentially
with time, demonstrating that the photon-created clusters quickly disappear
after the excitation and the system relaxes back to the paraelectric state.
Meanwhile, at temperature closer to $T_c$, more gentle decay rate is observed,
indicative of a critical slowing down of the relaxation time.
This is because with the decrease of temperature, the intensity of local
polarization is enhanced, and a long range correlation between dipole moments
is to be established as well, making the relaxation of photon-created
clusters slower and slower.

\section{Summary}

In conclusion, based on a coupled photon-phonon model, we conduct a theoretical
survey on the relaxation dynamics of nano-sized ferroelectric cluster
in the paraelectric phase of BaTiO$_3$, which is experimentally observed
as a real time correlation of speckle pattern between two soft X-ray
laser pulses.
The density matrix is calculated by a perturbative expansion up to the
fourth order terms, so as to characterize the time dependence of scattering
probability.
We show that this time dependence is determined by the relaxation dynamics
of photo-created ferroelectric cluster, which is manifested as a central
peak in the phonon spectral function.
Near the $T_c$, cluster excitation is stable with a long relaxation time.
While, at high temperature, it is suppressed by the thermal fluctuation,
resulting in a short relaxation time.
These features are in good agreement with Namikawa's experimental result.

\section*{Acknowledgments}
This work is supported by the Next Generation Supercomputer Project,
Nanoscience Program, MEXT, Japan.


\begin{thebibliography}{99}

\bibitem{po98} D. L. Polla and L. F. Francis,
  Processing and characterization of piezoelectric materials and integration
  into microelectromechanical systems,
  {\it Annu. Rev. Mater. Sci.} \textbf{28} (1998) 563-597.

\bibitem{ha71} J. Harada, J. D. Axe, and G. Shirane,
  Neutron-scattering study of soft modes in cubic BaTiO$_3$,
  {\it Phys. Rev. B} \textbf{4} (1971) 155-162.

\bibitem{mi76} R. Migoni, D. Bauer, and H. Bilz,
  Origin of Raman scattering and ferroelectricity in oxidic perovskites,
  {\it Phys. Rev. Lett.} \textbf{37} (1976) 1155-1158.

\bibitem{za03} B. Zalar, V. V. Laguta, and R. Blinc
  NMR evidence for the coexistence of order-disorder and displacive components
  in barium titanate,
  {\it Phys. Rev. Lett.} \textbf{90} (2003) 037601/1-4.

\bibitem{vo07} G. V\"{o}lkel and K. A. M\"{u}ller,
  Order-disorder phenomena in the low-temperature phase of BaTiO$_3$,
  {\it Phys. Rev. B} \textbf{76} (2007) 094105/1-8.

\bibitem{ya69} Y. Yamada, G. Shirane, and A. Linz,
  Study of critical fluctuations in BaTiO$_3$ by neutron scattering,
  {\it Phys. Rev.} \textbf{177} (1969) 848-857.

\bibitem{mu96} W. L. Mulvihill, K. Uchino, Z. Li and W. Cao,
  In-situ observation of the domain configuration during the phase transitions
  in barium titanate,
  {\it Phil. Mag. B} \textbf{74} (1996) 25-36.

\bibitem{pa98} G. K. H. Pang and K. Z. Baba-Kishi,
  Characterization of butterfly single crystals of BaTiO$_3$ by atomic
  force, optical and scanning electron microscopy techniques,
  {\it J. Phys. D} \textbf{31} (1998) 2846-2853.

\bibitem{go07} J. W. Goodman,
  {\it Speckle Phenomena in Optics: Theory and Applications}
  (Roberts and Company, 2007).

\bibitem{ta02} R. Z. Tai, K. Namikawa, M. Kishimoto, M. Tanaka, K. Sukegawa,
  N. Hasegawa, T. Kawachi, M. Kado, P. Lu, K. Nagashima, H. Daido,
  H. Maruyama, A. Sawada, M. Ando, and Y. Kato,
  Picosecond snapshot of the speckles from ferroelectric BaTiO$_3$ by
  means of X-ray lasers,
  {\it Phys. Rev. Lett.} \textbf{89} (2002) 257602/1-4.

\bibitem{ta04} R. Z. Tai, K. Namikawa, A. Sawada, M. Kishimoto, M. Tanaka,
  P. Lu, K. Nagashima, H. Maruyama, and M. Ando,
  Picosecond view of microscopic-scale polarization clusters in paraelectric
  BaTiO$_3$,
  {\it Phys. Rev. Lett.} \textbf{93} (2004) 087601/1-4.

\bibitem{na08} K. Namikawa, unpublished.

\bibitem{kr75} J. A. Krumhansl and J. R. Schrieffer,
  Dynamics and statistical mechanics of a one-dimensional model Hamiltonian
  for structural phase transitions,
  {\it Phys. Rev. B} \textbf{11} (1975) 3535-3545.

\bibitem{ru01} A. N. Rubtsov and T. Janssen,
  Quantum phase transitions in the discrete $\varphi^4$ model: The crossover
  between two types of transition,
  {\it Phys. Rev. B} \textbf{63} (2001) 172101/1-4.

\bibitem{na94} K. Nasu,
  Resonance enhancement of inelastic X-ray scatterings induced by strong
  visible lights,
  {\it J. Phys. Soc. Jpn.} \textbf{63} (1994) 2416-2427.

\bibitem{fe71} A. Fetter and J. Walecka,
  {\it Quantum Theory of Many-Particle Systems}
  (MacGrow Hill, New York, 1971).

\bibitem{ji04} K. Ji, H. Zheng, and K. Nasu,
  Path-integral theory for evolution of momentum-specified photoemission
  spectra from broad Gaussian to two-headed Lorentzian due to electron-phonon
  coupling,
  {\it Phys. Rev. B} \textbf{70} (2004) 085110/1-9.


\end{thebibliography}
\end{document}